\documentclass[%
 aip,
 jcp,%
 amsmath,amssymb,
 preprint,%
]{revtex4-2}

\usepackage{graphicx}% Include figure files
\usepackage{dcolumn}% Align table columns on decimal point
\usepackage{bm}% bold math
\usepackage{chemformula} % Formula subscripts using \ch{}
\usepackage[T1]{fontenc} % Use modern font encodings
\usepackage{amssymb}
\usepackage{amsmath}

\usepackage{multirow}
\usepackage{booktabs}
\usepackage{subcaption}
\RequirePackage[normalem]{ulem}
\RequirePackage{color}\definecolor{RED}{rgb}{1,0,0}\definecolor{BLUE}{rgb}{0,0,1}

\begin{document}

\preprint{Cite as: J. Chem. Phys. \textbf{151}, 234704 (2019); DOI: \url{https://doi.org/10.1063/1.5127513}}
\title{Electronic properties of Pb-I deficient lead halide perovskites}
%\title[Pb-I deficient perovskites]{Electronic properties of Pb-I deficient lead halide perovskites}% Force line breaks with \\
%\thanks{Footnote to title of article.}
\author{Chao Zheng}
\author{Oleg Rubel}
\email{rubelo@mcmaster.ca}
 \affiliation{Department of Materials Science and Engineering, McMaster University, Hamilton, ON L8S 4L8, Canada}%Lines break automatically or can be forced with \\
\author{Mika{\"e}l Kepenekian}%
\author{Xavier Rocquefelte}
\author{Claudine Katan}
\email{claudine.katan@univ-rennes1.fr}
\affiliation{ 
Univ Rennes, ENSCR, INSA Rennes, CNRS, ISCR (Institut des Sciences Chimiques de Rennes) - UMR 6226, F-35000 Rennes,
France%\\This line break forced with \textbackslash\textbackslash
}%

\date{\today}% It is always \today, today,


\begin{abstract}
The electronic structure evolution of deficient halide perovskites with a general formula $(A,A')_{1+x}M_{1-x}X_{3-x}$ was investigated using the density functional theory. The focus is placed on characterization of changes in the band gap, band alignment, effective mass, and optical properties of deficient perovskites at various concentrations of defects. We uncover unusual electronic properties of the defect corresponding to a $M\!-\!X$ vacancy filled with an $A'$ cation. This defect "repels" electrons and holes producing no trap states and, in moderate quantities ($x\le0.1$), does not hinder charge transport properties of the material. This behavior is rationalized using a confinement model and provides an additional insight to the defect tolerance of halide perovskites.
\end{abstract}

\maketitle

%\begin{quotation}
%The ``lead paragraph'' is encapsulated with the \LaTeX\ 
%\verb+quotation+ environment and is formatted as a single paragraph before the first %section heading. 
%(The \verb+quotation+ environment reverts to its usual meaning after the first %sectioning command.) 
%Note that numbered references are allowed in the lead paragraph.
%
%The lead paragraph will only be found in an article being prepared for the journal %\textit{Chaos}.
%\end{quotation}

\section{Introduction}\label{sec:Introduction}
%%%%%%%%%%%%%%%%%%%%
Recently, a new family of hybrid-halide perovskites called "deficient" perovskites (d-perovskites) and "hollow" perovskites  was reported \cite{Leblanc_ACIE_56_2017,Ke_SA_3_2017,Spanopoulos_JACS_140_2018,Ke_JACS_139_2017,Ke_AEL_3_2018,Leblanc_AAMI_11_2019}. Their structure is different from conventional $AMX_3$ perovskites where a relatively small organic cation $A^+$, such as methyl ammonium ($A^+=\mathrm{MA}^+$), is centered within an inorganic $MX_3^-$ cage formed by lead or tin halides. Deficient perovskites have missing $(M\!-\!X)^+$ units with corresponding voids filled by larger cations, such as hydroxyethylammonium ($A'^+=\mathrm{HEA}^+$), ethylenediammonium ($A'^+=\mathrm{EDA}^+$), or thioethylammonium ($A'^+=\mathrm{TEA}^+$). The defect ($d$) can be expressed using Kr{\"o}ger-Vink notations as
\begin{equation}\label{Eq:PbI-defect}
	d^{\times} = \text{v}''_\text{Pb} + \text{v}^{\bullet}_\text{I} + (A')^{\bullet}_\text{Pb--I}.
\end{equation}
The details of the defect reaction in Eq.~(\ref{Eq:PbI-defect}) indicates that $d$ consists of two intrinsic defects and one extrinsic defect. The superscript $\times$ indicates null charge for the current site. $\text{v}''_\text{Pb}$ is the first intrinsic defect which signifies the vacancy of Pb. The double prime $''$ denotes a negative charge $(-2)$ due to missing of a \ch{Pb^2+}. The second intrinsic defect is the vacancy of I $\text{v}^{\bullet}_\text{I}$, where the superscript $\bullet$ denotes a net single positive charge $(+1)$ due to missing of \ch{I-}. $(A')^{\bullet}_\text{Pb--I}$ is the extrinsic defect which demonstrates that $A'^+$ cation occupies the vacancies of Pb and I. This defect contributes a net single positive charge due to the presence of $A'^+$.

The general formula of d-perovskites is $(A,A')_{1+x}M_{1-x}X_{3-x}$ \cite{Leblanc_ACIE_56_2017}. The concentration $x$ of such structural defects can be quite large (the range of $0<x<0.2$ has been demonstrated experimentally \cite{Leblanc_ACIE_56_2017}) leading to a range of $3<X/M<3.5$ that captures a deviation from conventional stoichiometry of perovskites. The difference between deficient and hollow perovskites comes from an ordered vs stochastic arrangement of defects, respectively. Despite a large concentration of structural defects, the material retains its 3D architecture and can be considered as a transition between the 2D and 3D structures \cite{Katan_CR_119_2019}.

The most noticeable change in electronic properties of d-perovskites is the opening of an optical band gap. The band gap of d-\ch{MAPbI$_3$} with I/Pb ratio of 3.5 ($x=0.2$) increases from 1.55 to 2.15~eV \cite{Leblanc_ACIE_56_2017}. \citeauthor{Spanopoulos_JACS_140_2018}\cite{Spanopoulos_JACS_140_2018} also reported increase of the band gap from 1.52 to 1.85~eV for hollow \ch{MAPbI$_3$} perovskites (I/Pb ratio $= 3.6$) and from 1.25 to 1.42~eV for hollow \ch{MASnI3} perovskites (I/Sn ratio $= 3.6$).
Discrepancies in the experimental data can be attributed to different large cations (${\mathrm{HEA}}^+$ vs ${\mathrm{EDA}}^+$). One should also note that the d-perovskites have been reported to present a dual band gap marked by the existence of two shoulders in the UV/vis spectra [see Fig.~2(a) in Ref.~\citenum{Leblanc_ACIE_56_2017}] that leads to another uncertainty. Other noticeable properties of deficient/hollow perovskites are improved stability \cite{Ke_SA_3_2017,Spanopoulos_JACS_140_2018,Leblanc_AAMI_11_2019}, improved photoluminescence efficiency of hollow \ch{MASnI3} \cite{Ke_JACS_139_2017}, and improved performance of Sn-based perovskite solar cells \cite{Ke_AEL_3_2018}.

%	VERSION BEFORE MK EDIT
%Discrepancies in the experimental data can be attributed to different large cations (${\mathrm{HEA}}^+$ vs ${\mathrm{EDA}}^+$) as well as to a dual nature of the band gap in d-perovskites \cite{Leblanc_ACIE_56_2017} leading to another uncertainty.
%%
%\CZadd{Dual band gap indicates for a material system, there are two emissions in photoluminescence spectroscopy studies, or two shoulders in UV/Vis spectra.}
%%
%Especially, for d-perovskite, \citeauthor{Leblanc_ACIE_56_2017} observed the presence of two shoulders in UV/Vis spectra [see Fig.~2(a) in Ref. \citenum{Leblanc_ACIE_56_2017}] and reported the corresponding second band gap was 2.15~eV.
%%	THIS SHOULD NOT BE SAID HERE.
%\CZadd{The authors did not explain the cause of this dual band gap problem in these materials. In the work, we investigated the d-perovskite structures using first-principles method and attributed the dual band gap to a confinement effect of defect sites.}
%%
%Other noticeable properties of deficient/hollow perovskites are improved stability \cite{Ke_SA_3_2017,Spanopoulos_JACS_140_2018,Leblanc_AAMI_11_2019}, improved photoluminescence efficiency of hollow \ch{MASnI3} \cite{Ke_JACS_139_2017}, and improved performance of Sn-based perovskite solar cells \cite{Ke_AEL_3_2018}.

\begin{figure}[h]
  \includegraphics[scale=0.35]{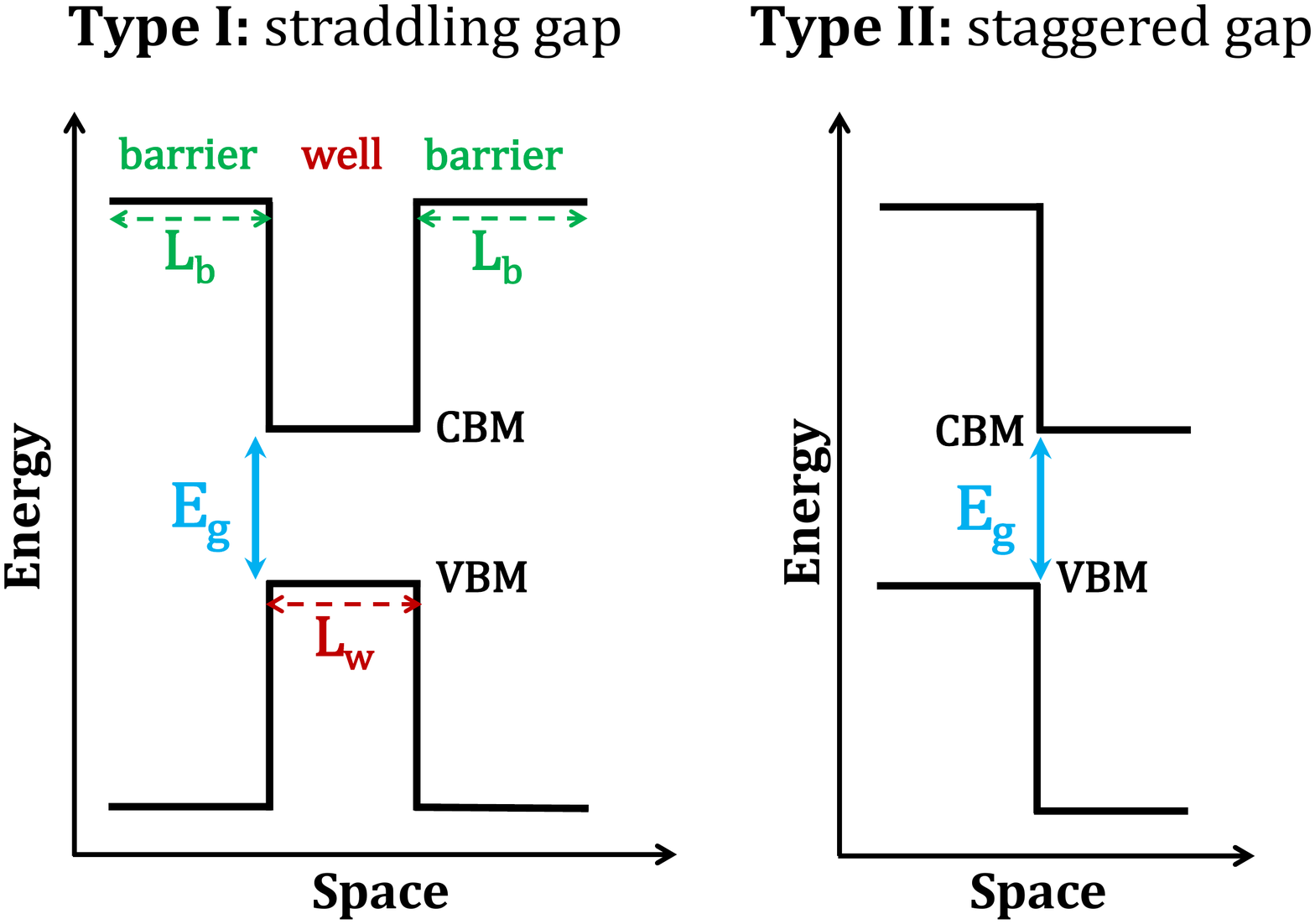}
  \caption{Band offsets at a semiconductor heterojunction illustrated for two types of band alignment . In type I, the conduction band of the well is lower than that of the barrier. Oppositely, the valence band of the well lies higher than that of the barrier. In type II, both the conduction and valence bands of one of the two materials building the heterojunction are located below those of the second material.}
  \label{Fig:Type-I-II}
\end{figure}

If the issue of defects in halide perovskites and their impact on photovoltaic or light emission performances has been intensively investigated (see for instance Refs.~\citenum{Ran_CSR_2018, Wang_npjfe_2018} and references herein),
theoretical studies of d-perovskites are scarce and limited to Refs. \citenum{Spanopoulos_JACS_140_2018,Ke_AEL_3_2018,Leblanc_AAMI_11_2019}. \textit{Ab initio} electronic structure calculations corroborate experimentally observed opening of the band gap; the band structure remains direct even at a high concentration of defects. It was also noticed that bands become less dispersed in hollow and d-perovskites \cite{Spanopoulos_JACS_140_2018,Leblanc_AAMI_11_2019}. The valence band dispersion in a lateral direction (i.e., perpendicular to defect tunnels) practically vanished in d-\ch{FAPbI$_3$} [Fig.~5(b) in Ref. \citenum{Leblanc_AAMI_11_2019}] at a high concentration of defects ($x=0.2$) that corresponds to formation of a continuous chain of defects (tunnels). \citeauthor{Leblanc_AAMI_11_2019}\cite{Leblanc_AAMI_11_2019} also pointed that the band gap opening with increasing $x$ occurs mostly due to a shift of the conduction band edge (CBE) upward in energy, while the valence band edge (VBE) shifts down by approximately 0.1~eV. Similar to semiconductor heterojunctions induced band offset, defects will introduce energy misalignment. Fig.~\ref{Fig:Type-I-II} illustrated two different types of band offsets, namely the type I band alignment or straddling gap and the type II band alignment or staggered gap. Identification of the type of band alignment is critical for investigating carrier transport properties in semiconductors.
%	THIS SHOULD NOT BE SAID HERE.
%In the work, we investigate the band alignment of d-perovskite and discuss the related electronic properties.}

Here, we use first-principle calculations to investigate the electronic structure evolution of d-\ch{MAPbI$_3$}. Two deficient structures are selected with the ratio $\text{I/Pb} = 3.22$ and 3.5 to represent moderate ($x=0.1$) and high density of defects ($x=0.2$), respectively. We quantify changes in the band gap, band alignment, effective mass, and effective band structure of d-\ch{MAPbI$_3$} based on the structural model from Ref. \citenum{Leblanc_ACIE_56_2017}. We anticipate the results to provide further insight into understanding a defect tolerance of perovskites \cite{Kim_JPCL_5_2014,Taufique_JCP_149_2018}.

\section{\label{sec:level1}Method}
%%%%%%%%%%%%%%%%%%%%
Density functional theory \cite{Hohenberg_PR_136_1964,Kohn_PR_140_1965} (DFT) calculations were performed using the Vienna ab initio simulation package \cite{Kresse_PRB_54_1996} (VASP) and projector augmented-wave potentials \cite{Kresse_PRB_54_1996,Kresse_PRB_59_1999,Blochl_PRB_50_1994}. A Perdew-Burke-Ernzerhof \cite{Perdew_PRL_77_1996} (PBE) gradient approximation for the exchange-correlation functional was chosen in combination with the \citeauthor{Grimme_JCP_132_2010}\cite{Grimme_JCP_132_2010} (D3) correction to capture long-range van der Waals interactions.

\begin{figure}[h]
  \includegraphics{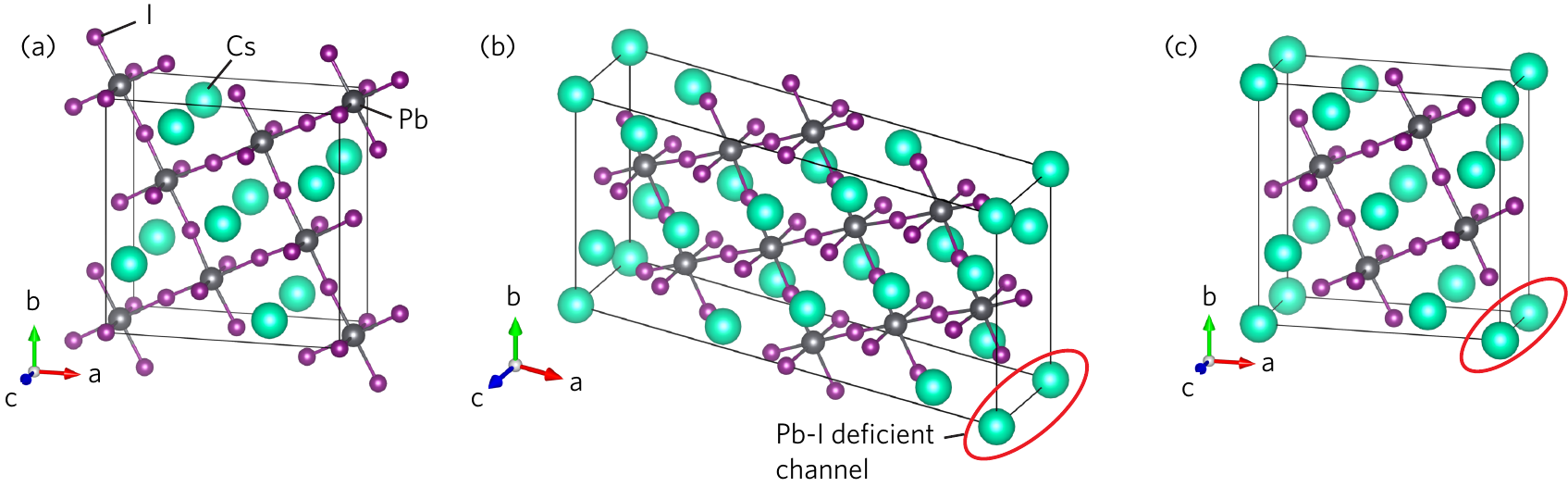}
  \caption{(a) Stoichiometric perovskite structure without deficiencies, (b) deficient perovskite with I/Pb ratio of 3.22 ($x=0.1$), (c) I/Pb ratio of 3.5 ($x=0.2$). Cesium atoms are used instead of MA cations. Defects are arranged in continuous channels along $c$ axis.}
  \label{Fig:Structures}
\end{figure}

The \ch{Pb-I} framework was kept fixed at experimental positions and lattice parameters \cite{Leblanc_ACIE_56_2017}. This implies that the same lattice constant and atomic distortions apply to all compositions, including the stoichiometric phase of \ch{MAPbI$_3$} [see supporting information (SI), Table~SI]. The experiment \cite{Leblanc_ACIE_56_2017} indicates only a small increase of the lattice parameters for d-\ch{MAPbI$_3$}. Cesium atoms were used instead of MA and HEA cations (Fig.~\ref{Fig:Structures}) to reduce computational efforts. This approach is justified as long as we preserve experimental positions for the lead-iodide framework, which determines the band gap and dispersion near to the band edges. In order to ensure that the conclusions drawn in this paper are not artifacts of the approximation made regarding the structure of d-perovskites, we investigate a structure where ${\mathrm{HEA}}^+$ organic cation is substituted in place of \ch{(Pb-I)^+} unit in tetragonal \ch{MAPbI$_3$}. Calculations for d-\ch{MAPbI$_3$} with HEA (without Cs and including relaxation of atomic positions for the entire structure) are presented in SI (Figs.~S1 and S2). Overall trends are the same (including the type-I band alignment, see Fig.~\ref{Fig:Type-I-II}), but effects are weaker (smaller band gap opening, smaller band discontinuities, and not so large effective mass enhancement).

When relaxing positions of Cs atoms, the cutoff energy $E_{\mathrm{cut}}$ for the plane-wave expansion was increased by 25\% above a recommended maximum value in pseudopotential files (VASP tag PREC=High, $E_{\mathrm{cut}}=275$~eV). In band structure calculations, $E_{\mathrm{cut}}$ corresponded to the recommended maximum value in pseudopotential files (VASP tag PREC=Normal,  $E_{\mathrm{cut}}=220$~eV). The absolute change in the band gap as a result of different $E_{\mathrm{cut}}$ is marginal (not more than 5~meV). The \citeauthor{Monkhorst_PRB_13_1976}\cite{Monkhorst_PRB_13_1976} shifted $k$ mesh was used for the structural relaxation (Cs atoms only), while the $\Gamma$-centred mesh was selected for calculations of the band gap and optical properties (see SI, Table~SI). Linear optical properties were calculated within an independent particle approximation ignoring local field effects. The spin-orbit coupling (SOC) was taken into account. The number of bands was increased twice from its default value when calculating optical properties. Other relevant parameters are summarized in Table~SI (see SI).

Effective masses were determined at the PBE+SOC level by fitting the energy dispersion $E(k)$ to a second order polynomial within the range of $\Delta k=\pm 0.003$~{\AA}$^{-1}$ near to the band edge sampled with $9-11$ intermediate $k$ points.

Band structure unfolding and Bloch character calculations were done with the fold2Bloch code \cite{Rubel_PRB_90_2014} following the method outlined in Ref. \citenum{Wang_PRL_80_1998}. The code computes spectral weights $w_n(\bm{k})$ in the range $0\!-\!1$ for each $n$'s energy eigenstate to reflect its Bloch $\bm{k}$ character. The number of spectral weights computed per eigenstate corresponds to the multiplicity of the supercell. The approach to the calculation of spectral weights is based on remapping the supercell reciprocal space with a mesh that is compatible with the translational symmetry of a primitive cell. Computed Bloch spectral weights for each eigenstate in a supercell calculation fulfill the normalization $\sum_{\bm{k}} w_n(\bm{k})=1$, which implies that the cumulative probability of finding an electron in any of multiple $\bm{k}$'s states adds up to one. Further details of the method and implementation can be found elsewhere \cite{Wang_PRL_80_1998,Rubel_PRB_90_2014}.

Vesta \cite{Momma_JAC_44_2011} was used for visualization of atomic structures. Structure files used in this work can be accessed at the Cambridge Crystallographic Data Center (CCDC) under deposition No. 1962341-1962347 and 1962349.
%%%%%%%%%%%%%%%%%%%%
\section{\label{sec:level1}Results and Discussion}
%%%%%%%%%%%%%%%%%%%%%
\subsection{\label{sec:level2}Structure of d-perovskites  and defect formation enthalpy}
It is worth commenting on the structure of d-perovskites before going into discussion of their electronic properties. Here we adopt the structure proposed in Ref. \citenum{Leblanc_ACIE_56_2017}, where the lattice parameters and coordinates of \ch{Pb-I} framework are specified. Fig.~\ref{Fig:Structures} shows the stoichiometric structure as well as two d-perovskite structures that differ by the I/Pb ratio (3.22 and 3.5) corresponding to the defect concentration of $x=0.1$ and 0.2, respectively. The closer the ratio to its stoichiometric value of 3, the less deficient is the perovskite structure. The ratio of $\text{I/Pb}=3.5$ is at the upper limit of deficiency attained experimentally \cite{Leblanc_ACIE_56_2017}, if not hollow. In the sections that follows, Cs atoms are used instead of organic cations (MA or HEA) for computational performance reasons (see also the case study with HEA in Method section).

The self-compensation defect formed by $A$ cation substituted for $M\!-\!X$ unit might also be unintentionally present in smaller concentrations in stoichiometric $AMX_3$ perovskite structures. To evaluate a formation energy of the corresponding defect [Eq. (1)] we built a $2\times2\times2$ supercell of d-\ch{MAPbI$_3$} with and without defect as shown in Fig.~S1 (see SI). The defect structure corresponds to the larger ${\mathrm{HEA}}^+$ cation substituted in place of a \ch{Pb-I} unit of \ch{MAPbI$_3$}. The corresponding reaction can be expressed as
\begin{equation}\label{Eq:d-perovsk-supercell-reaction}
	\ch{MA8Pb8I24} + \ch{HEAI} \rightarrow \ch{MA8(HEA)Pb7I23} + \ch{PbI2}.
\end{equation}
Giving the size of the supercell, it corresponds to a defect concentration of $x=0.125$. We assign the DFT total energy difference between reactants and products to the defect formation enthalpy
\begin{equation}\label{Eq:def-energy}
	\Delta H_d = E_{\ch{MA8Pb8I24}} + E_{\ch{PbI2}} - \left( E_{\ch{MA8(HEA)Pb7I23}} - E_{\ch{HEAI}} \right).
\end{equation}

The calculations yield $\Delta H_d = 0.78$~eV. This result can be compared with another common Schottky defect in perovskites, namely \ch{PbI2} vacancy. The later has a formation energy of about 0.2~eV \cite{Saidaminov_NE_3_2018} suggesting that the defect described by Eq. (1) is less likely to form spontaneously.

\subsection{\label{sec:level2}Band gap opening, band alignment, dual band gap and optical properties}
There is a significant opening of the band gap in the deficient structures relative to the reference (stoichiometric) perovskite. Computed band gaps are listed in Table~\ref{tbl:Gaps}. The band gap opening is more prominent when relativistic effects are included. The calculated band gap increment (0.74~eV with SOC and 0.36~eV without SOC) is within the range of the experimental band gap opening quoted in  Sec.~\ref{sec:Introduction}.

\begin{figure*}[h]
  \includegraphics{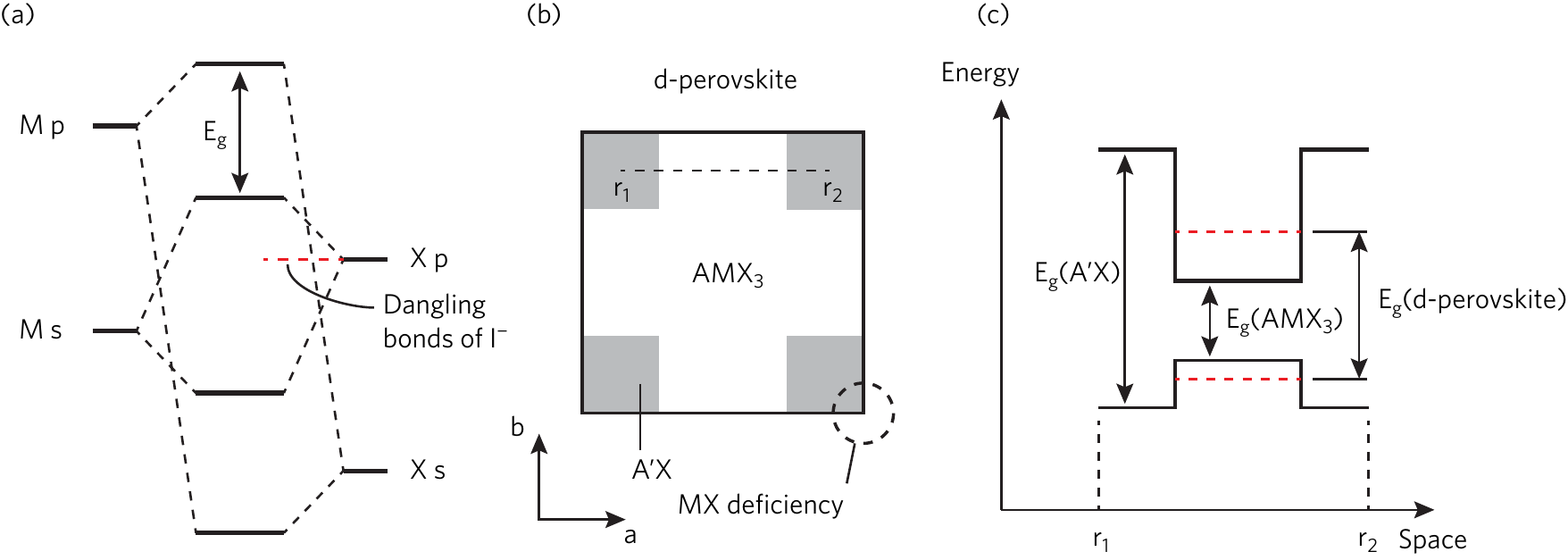}
  \caption{(a) Simplified orbital energy diagram of $AMX_3$. Only states of the $MX_3^-$ lattice are shown. SOC effects as well as $s\!-\!s$ and $p\!-\!p$ hybridizations are omitted. (b) Schematic structure of d-perovskite. Gray regions correspond to the local $A'X$ chemical environment. (c) Band diagram plotted along the $r_1 - r_2$ path. Electronic states of d-perovskite are derived from $AMX_3$ confined between regions of $A'X$. Since holes are less confined than electrons, the corresponding confinement energies are ordered as $\delta_\mathrm{c}>\delta_\mathrm{v}$.}
  \label{Fig:Confinement}
\end{figure*}

It is quite unusual that defects lead to opening of the band gap. It implies that the defect described in Eq.~(\ref{Eq:PbI-defect}) does not create states within the band gap. The \ch{Pb-I} vacancy introduces five undercoordinated I atoms and one Pb atom. A simplified orbital energy diagram of halide perovskites in Fig.~\ref{Fig:Confinement}(a) is intended to rationalize this result. Since both top of the valence band and bottom of the conduction band are formed by antibonding orbitals \cite{Umebayashi_PRB_67_2003,Tao_NC_10_2019}, compensated dangling bonds of \ch{I^-} are expected to appear \textit{within} the bulk of valence band states as confirmed by the analysis of atom-specific contribution to energy eigenstates (see SI, Fig.~S3).

Opening of the band gap can be viewed as a confinement effect. Defects form local regions of $A'X$ with a wide band gap and confine electronic states within the perovskite region as shown in Fig.~\ref{Fig:Confinement}(b,c). The higher the concentration of defects, the smaller the distance between $A'X$ regions leading to a higher confinement energy. Since it is a confinement effect, we would expect it to be operational at a high enough concentration of defects when the distance between them becomes comparable to the lattice spacing.

\begin{figure}[h]
  \includegraphics{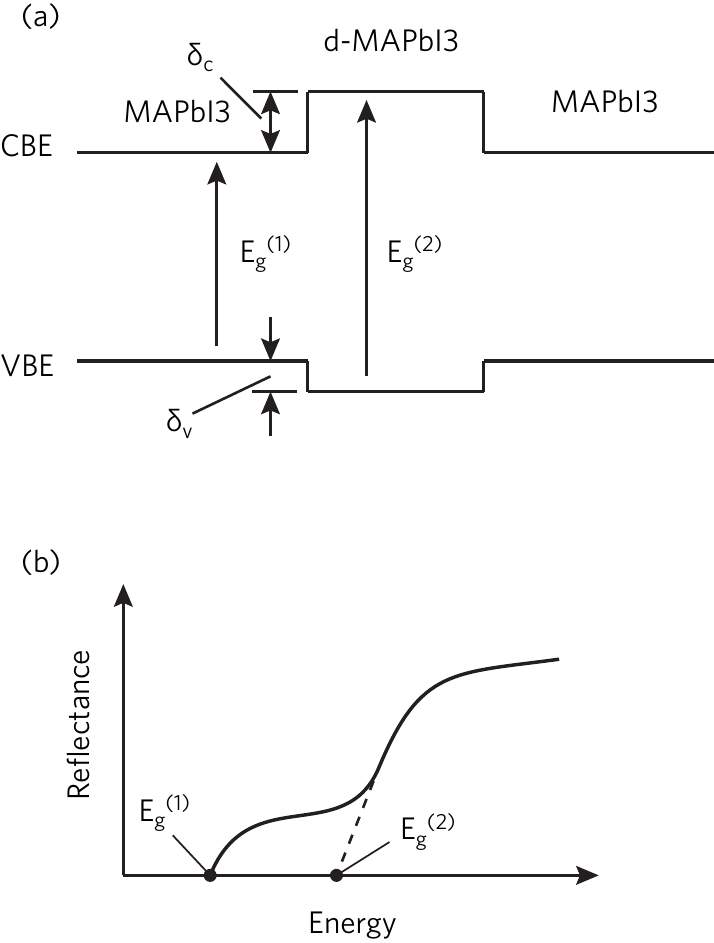}
  \caption{(a) Band alignment diagram that shows a d-perovskite region embedded in the stoichiometric \ch{MAPbI$_3$}. (b) Photoreflectance spectrum (schematic) that shows two apparent band gaps, that originate from two optical transitions $E_\mathrm{g}^{(1)}$ and $E_\mathrm{g}^{(2)}$ shown on panel (a).}
  \label{Fig:Alignment}
\end{figure}

It is interesting to see whether this type of defects will act as trap for charge carriers. The greater band gap of d-perovskites alone does not provide an answer to this question, as there are different types of band alignment possible (namely type I or II, Fig.~\ref{Fig:Type-I-II}). The band alignment calculations were performed using core Cs-$5s$ states as an energy reference to relate position of the valence and conduction bands in different calculations. (See Refs. \citenum{Wei_PRB_60_1999,Gupta_JAP_126_2019} for details of this method as well as discussion of its accuracy and validation.) Our calculations suggest an energy misalignment of approximately $\delta_\mathrm{v}=0.1$~eV (Table~\ref{tbl:Gaps}) between the VBE of stoichiometric and d-perovskite with VBE of d-perovskite being lower in energy as shown in Fig.~\ref{Fig:Alignment}(a). This result is consistent with the recent data on d-\ch{FAPbI$_3$} \cite{Leblanc_AAMI_11_2019}. Most of the band gap difference between stoichiometric and d-perovskite comes from the CBE offset $\delta_\mathrm{c}$ (Table~\ref{tbl:Gaps}). This result indicates a type-I band alignment between stoichiometric and d-perovskites. Thus, \ch{Pb-I} deficient regions should pose no threat for trapping of charge carriers. It is an interesting result, since defects are generally expected to create either shallow or deep traps within the band gap, which is not the case here.

The disparity between the CBE and VBE band offset energies ($\delta_\mathrm{c}$ and $\delta_\mathrm{v}$) can be attributed to a much greater confinement energy for electrons than holes in the $A'X/AMX_3$ heterostructure shown in Fig.~\ref{Fig:Confinement}(b,c). The diagram was constructed based on experimental data for an electron affinity of 0.63~eV \cite{Miller_JCP_85_1986} and an ionization energy of 7.1~eV \cite{Benson_JPBAMP_20_1987} for CsI as well as an electron affinity of 4.4 eV and an ionization energy of 5.9 eV for \ch{MAPbI$_3$} \cite{Tao_NC_10_2019}.

The band diagram in Fig.~\ref{Fig:Alignment}(a) can now be used to explain an apparent dual band gap in reflectance measurements [see Fig.~2(a) in Ref. \citenum{Leblanc_ACIE_56_2017} also shown schematically in Fig.~\ref{Fig:Alignment}(b)]. The dual band gap character becomes more prominent with increasing deficiency of the perovskite. The first and second experimental gaps amount to $E_\mathrm{g}^{(1)}\approx1.6$~eV and $E_\mathrm{g}^{(2)}\approx2.0-2.1$~eV, respectively. 

\begin{figure}[h]
  \includegraphics{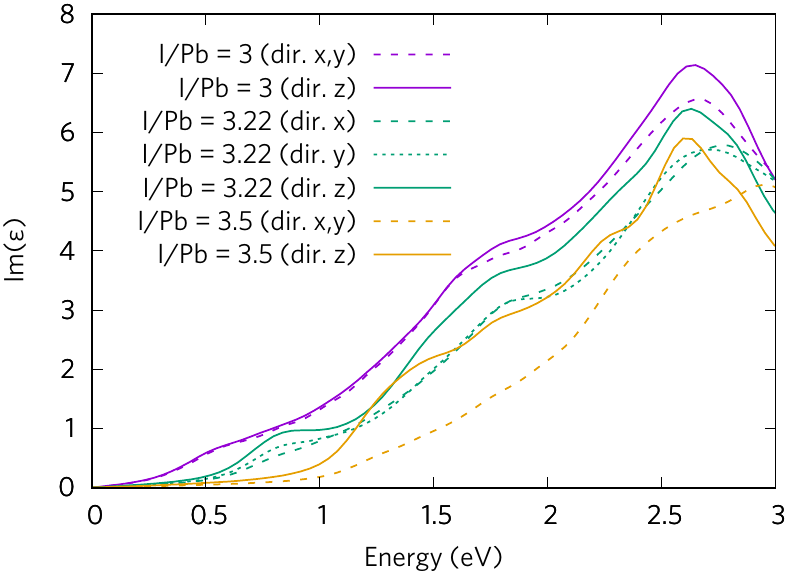}
  \caption{Frequency-dependent imaginary part of the dielectric function for a stoichiometric perovskite structure without deficiencies ($x=0$, violet) and d-perovskites with I/Pb ratio of 3.22 and 3.5 ($x=0.1$, green, and 0.2, orange, respectively). Only diagonal components of the dielectric tensor are presented ($\epsilon_{xx}$, $\epsilon_{yy}$ and $\epsilon_{zz}$); off-diagonals components are close to zero. The Cartesian directions correspond to the coordinate system in Fig.~\ref{Fig:Structures}.}
  \label{Fig:Optics}
\end{figure}

To further inspect the possible origin of the dual band gap, we compute the optical response of d-perovskites. Calculations of optical properties (Fig.~\ref{Fig:Optics}) indicate an anisotropy of the dielectric function in d-perovskites. The anisotropy becomes more prominent as the deficiency increases. The onset of absorption in d-perovskites is shifted to higher energies following the band gap trend and in accord with experimental observations \cite{Leblanc_ACIE_56_2017,Spanopoulos_JACS_140_2018}. The calculated imaginary part of the dielectric function (Fig.~\ref{Fig:Optics}) shows a clear blue shift, but no features that can be interpreted as the second gap. The monotonic behavior could be attributed to the growing density of states as the energy moves away from band edges into the bands (see SI, Fig.~S4). Thus, to rationalize experimental observation, one needs to assume a coexistence of stoichiometric \ch{MAPbI$_3$} regions that contribute to $E_\mathrm{g}^{(1)}$ and \ch{Pb-I} deficient regions with higher optical transition energies at $E_\mathrm{g}^{(2)}$ as shown in Fig.~\ref{Fig:Alignment}(a).

\subsection{\label{sec:level2}Carrier transport properties}
\begin{figure*}[h]
  \includegraphics{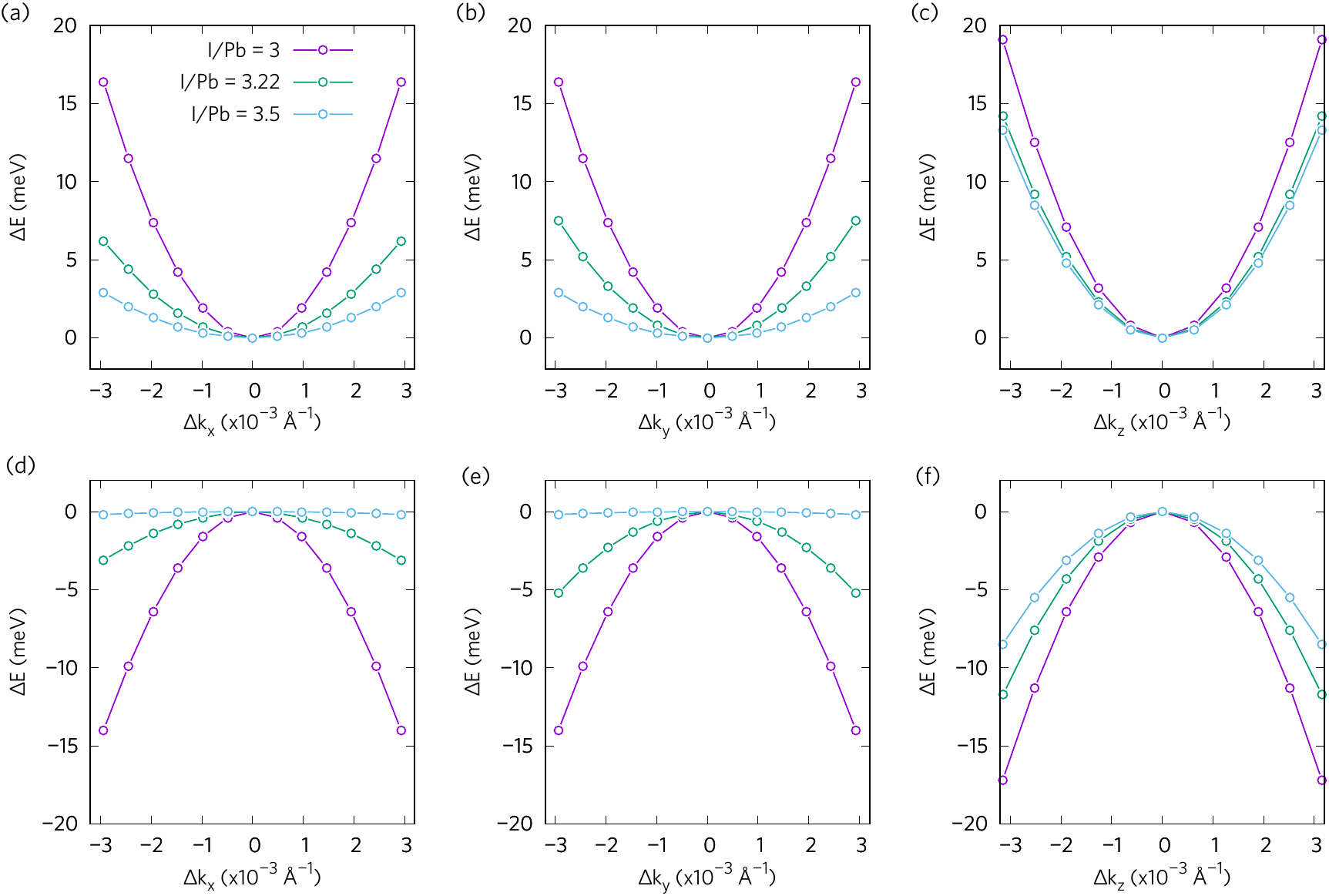}
  \caption{Band dispersion along $x$, $y$, and $z$ direction in reciprocal space in the vicinity of (a-c) CBE and (d-f) VBE for a stoichiometric perovskite structure without deficiencies ($\text{I/Pb}=3$, $x=0$) and deficient perovskites with I/Pb ratios of 3.22 and 3.5 ($x=0.1$ and 0.2,\ respectively). Bands become less dispersive with increasing the deficiency $x$ in particular in the lateral directions ($x$ and $y$).}
  \label{Fig:Dispersion}
\end{figure*}

Since d-perovskites are considered as a potential absorber material in solar cells,\cite{Leblanc_ACIE_56_2017,Ke_AEL_3_2018} it is interesting to explore how deficiencies affect intrinsic transport characteristics of this material. The band dispersion near CBE and VBE is shown in Fig.~\ref{Fig:Dispersion}. As a general trend, d-perovskites show less dispersive bands, which was also noticed previously \cite{Spanopoulos_JACS_140_2018}. The band dispersion is mostly affected within the $x\!-\!y$ plane. It is the plane perpendicular to direction of the tunnel formed after removing the \ch{Pb-I} unit [Fig.~\ref{Fig:Structures}(c)]. At high concentration of defects, the dispersion remains only along $z$ axis that electronically resembles 1D structures.

Effective mass calculations (Table~\ref{tbl:Mstar}) corroborate conclusions drawn from the band dispersion data in Fig.~\ref{Fig:Dispersion}. The effective mass magnitude $|m^*/m_e|$ in the stoichiometric perovskite is of the order of 0.09 for electrons and holes. Our calculated effective masses are about twice less that previous calculations for \ch{CsPbI$_3$}, $|m^*/m_e|=0.15-0.20$ \cite{Hendon_JMCA_3_2015,Jong_PRB_98_2018} and for \ch{MAPbI$_3$} $|m^*/m_e|=0.13-0.22$ \cite{Menendez-Proupin_PRB_90_2014}, and are also underestimated when compared to more accurate results based on many body perturbation theory within the GW approximation \cite{Umari_SR_4_2014,Filip_JPCC_119_2015}. It is expected that the bare DFT-PBE will underestimates effective masses \cite{Kim_PRB_82_2010} due to the band gap error. Meanwhile, our computed reduced effective masses ($0.08-0.09$) compare quite well to the experimental exciton reduced mass of \ch{MAPbI$_3$} $0.104|m_e|$ \cite{Miyata_NP_11_2015}. In d-perovskites, the effective masses of electrons and holes increases as the material becomes more defective. At the intermediate concentration of defects (the I/Pb ratio of 3.22, $x=0.1$), the increase is moderate (by a factor of 4 at most). The effective mass becomes anisotropic. The lightest effective mass corresponds to the band dispersion along channels. The charge transport (especially holes) becomes effectively 1D when the concentration of defects (\ch{Pb-I} vacancies) approached the upper limit ($\text{I/Pb}=3.5$, $x=0.2$) and defects form a hollow tunnel along $z$ axis.

\subsection{\label{sec:level2}Band unfolding of d-perovskite}
\begin{figure}[h]
  \includegraphics{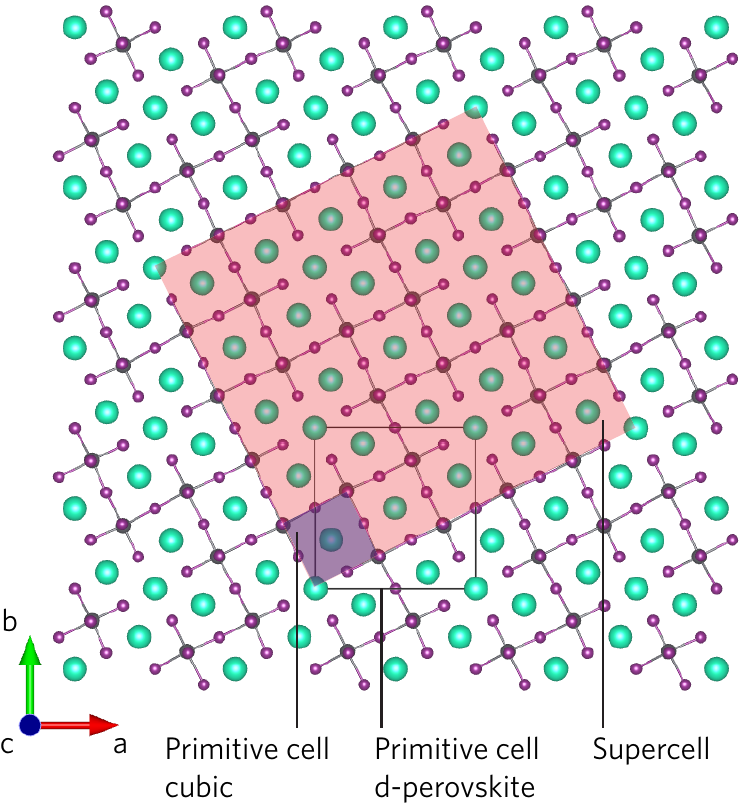}
  \caption{Relation between the d-perovskite unit cell (black solid line), the cubic primitive cell (blue area), and the supercell (pink area).}
  \label{Fig:SCell}
\end{figure}

To gain further insight into changes of the band structure introduced by \ch{Pb-I} deficiency we employ a band unfolding technique. It requires the supercell to have lattice vectors aligned with lattice vectors of the primitive cell. Here we use the quasi-cubic cell (blue area on Fig.~\ref{Fig:SCell}) as a primitive unit for projecting an unfolded band structure. We created a supercell in such a way that its lattice vectors are aligned with those of the primitive cubic cell while preserving the periodicity of the deficient perovskite structure at the same time (Fig.~\ref{Fig:SCell}). The smallest supercell that fulfills these conditions is shown as a pink area in Fig.~\ref{Fig:SCell}. Its multiplicity factor is $5\times5\times2$ relative to the cubic primitive cell.

\begin{figure*}[h]
  \includegraphics{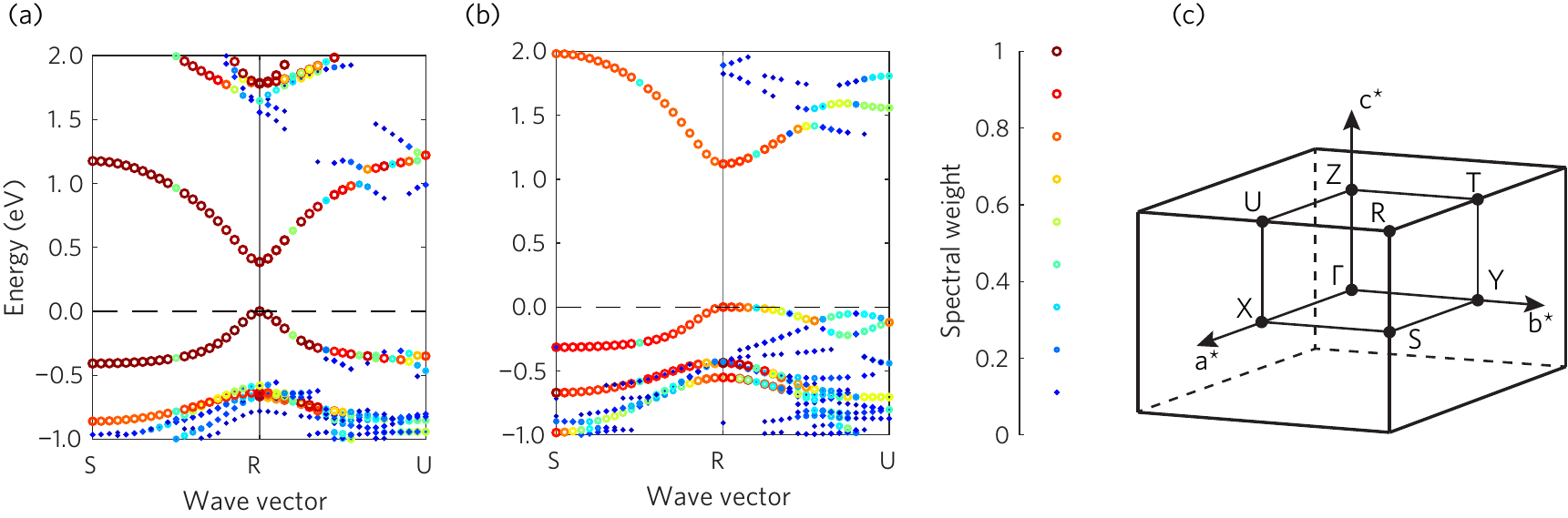}
  \caption{Effective band structure of (a) stoichiometric and (b) d-perovskite ($\text{I/Pb}=3.5$, $x=0.2$). The color and size of symbols correspond to the Bloch spectral weight. Points with the spectral weight less than 0.05 are not shown. (c) Brillouin zone of an orthorhombic lattice. The Bloch character is well preserved (greater than 70\%) in d-perovskites despite of less dispersive bands.}
  \label{Fig:Unfolding}
\end{figure*}

The unfolded band structures are shown in Fig.~\ref{Fig:Unfolding}. A band structure of the stoichiometric phase was also unfolded for comparison. It shows a direct band gap at $R$ point [Fig.~\ref{Fig:Unfolding}(a)] as expected from previous studies of 3D lead-halide perovskites \cite{Umebayashi_PRB_67_2003,Koutselas_JPCM_8_1996}. The Bloch character of electronic states near to the band edges of the stoichiometric perovskite is almost 100\%. In the d-perovskite with high density of defects ($\mathrm{I/Pb}=3.5$, $x=0.2$), $70-80$\% of the Bloch character is preserved [Fig.~\ref{Fig:Unfolding}(b)]. The band structure remains direct at $R$ point. The band dispersion characteristics reflect the above discussion of effective masses. A well-preserved Bloch character near $R$ point implies that states are not localized despite the presence of structural defects. 
This is further corroborated when using inverse participation ratio \cite{Pashartis_PRA_7_2017} of electronic eigenstates as a measure of localization (see SI for details). In fact, analysis of the inverse participation ratio (see SI, Fig.~S5) did not reveal any localized states near to the band edges either. This unusual behavior of d-perovskites is another manifestation of a defect tolerance inherent to hybrid halide perovskites \cite{Kim_JPCL_5_2014,Yin_APL_104_2014}.

%%Moved to method %% Calculations for d-\ch{MAPbI3} with HEA (without Cs and including relaxation of atomic positions for the entire structure) are presented in SI (Figs.~S1 and S2). Overall trends are the same (including the type-I band alignment), but effects are weaker (smaller band gap opening, smaller band discontinuities, and not so large effective mass enhancement).

%% Moved to conclusion %% Finally, we would like to comment on the electronic structure dimensionality of d-perovskites. Deficient perovskites with a moderate density of defects ($\mathrm{I/Pb}=3.22$, $x=0.1$) behave as a 3D material with anisotropic effective masses (Table~\ref{tbl:Mstar}). As the density of defects increases ($\mathrm{I/Pb}=3.5$, $x=0.2$), hollow regions connect into tunnels along $z$ axis, which remains the only direction for efficient band dispersion leading to a 1D electronic structure. Also, the band gap opening of d-perovskites exceeds that of 2D perovskites. This statement builds on the fact that a PL emission wavelength of 2D 1~ML \ch{MAPbI3} perovskite is centered at 720~nm \cite{Liu_AN_10_2016}, while d-\ch{MAPbI3} have the secondary gap at shorter wavelengths (about 600~nm \cite{Leblanc_ACIE_56_2017}).
%%%%%%%%%%%%%%%%%%%%%
\section{\label{sec:level1}Conclusion}
%%%%%%%%%%%%%%%%%%%%%
The electronic structure evolution of d-perovskites with a general formula $(A,A')_{1+x}M_{1-x}X_{3-x}$ was investigated using the density functional theory for an intermediate and a high concentration of defects ($x=0.1$ and 0.2, respectively). The formation enthalpy of \ch{Pb-I} deficient defects filled with a larger ${\mathrm{HEA}}^+$ cation for charge compensation is about 0.8~eV, which is relatively large when compared to other native defects in halide perovskites. Our calculations reproduce the opening of a band gap also observed in previous theoretical studies. The band gap opening is achieved mostly due to moving the conduction band edge upward in energy. In addition, the valence band edge shifts slightly toward lower energies by approximately 0.1~eV. This leads to the formation of a type-I band alignment between d-perovskites and the stoichiometric phase. As a result, \ch{Pb-I} deficient regions do not act as traps for charge carriers. This result is rationalized via a confinement model where the defect is viewed locally as a wide band gap region of $A'X$.

Presence of this type of defects in moderate quantities ($x\le0.1$) should not hinder the ability of perovskite material to function as a good solar cell absorber. Based on effective mass calculations, we expect a reduced mobility of electrons and holes in d-\ch{MAPbI$_3$} ($x=0.1$) by a factor of 4 at most. However, at high concentration of defects ($x=0.2$), the charge transport ceases within a lateral plane due to an extremely heavy effective mass of holes thereby posing an issue for extracting photogenerated charge carriers. Interestingly, no localized states were detected near to the band edges of d-perovskites. It is attributed to the fact that both conduction and valence band edges are derived from antibonding orbitals.

Investigation of the optical properties of d-perovskites reveal no anomalies in absorption spectra. The absorption spectrum increases monotonously up to a photon energy of about 2.5~eV.  Thus, dual band gap features observed experimentally at a high concentration of defects ($x>0.15$) are attributed to coexistence of deficient and stoichiometric regions.

Finally, we would like to comment on the electronic structure dimensionality of d-perovskites. Deficient perovskites with a moderate density of defects ($\mathrm{I/Pb}=3.22$, $x=0.1$) behave as a 3D material with anisotropic effective masses. As the density of defects increases ($\mathrm{I/Pb}=3.5$, $x=0.2$), hollow regions connect into tunnels along the $z$ axis, which remains the only direction for efficient band dispersion leading to a 1D electronic structure. Also, the band gap opening of d-perovskites exceeds that of 2D perovskites. This statement builds on the fact that a PL emission wavelength of 2D 1~ML \ch{MAPbI$_3$} perovskite is centered at 720~nm \cite{Liu_AN_10_2016}, while d-\ch{MAPbI$_3$} have the secondary gap at shorter wavelengths (about 600~nm \cite{Leblanc_ACIE_56_2017}).

\section*{Supplementary material}

See the supplementary material at \url{https://doi.org/10.1063/1.5127513} for parameters used in electronic structure calculations; structure and band dispersion results for d-\ch{MAPbI3}; density of states, partial density of states, and the inverse participation ratio analysis. References cited therein are \cite{Zheng_PRM_2_2018,Pashartis_PRA_7_2017,Rubel_git_VASPtools}.

%%%%%%%%%%%%%%%%%%%%%
\begin{acknowledgments}
Authors are indebted to Prof. Jacky Even from INSA Rennes for fruitful discussions. C.Z. and O.R. acknowledge funding provided by the Natural Sciences and Engineering Research Council (NSERC) of Canada under the Discovery Grant Program RGPIN-2015-04518. O.R. would like to acknowledge a mobility grant ``Materials for green energy" provided by the McMaster University in the framework of International Micro-Fund Initiatives. C.K. and M.K. acknowledge support from Agence Nationale pour la Recherche (MORELESS project). Calculations were performed using a Compute Canada infrastructure supported by the Canada Foundation for Innovation under the John R. Evans Leaders Fund program. C.Z. acknowledges the mobility grant provided by Metropole Rennes.
\end{acknowledgments}
%%%%%%%%%%%%%%%%%%%%%%
%Dedication
\begin{quotation}
This paper is dedicated to Dr. Jean-Fran\c{c}ois Halet at the occasion of his 60th birthday. It is also a tribute to the memory of Prof. George Papavassiliou.
\end{quotation}
%\nocite{*}
\bibliography{manuscript}% Produces the bibliography via BibTeX.
%%%%%%%%%%%%%%%%%%%%

%%%%%%%%%%%%%%%%%%%%%
\clearpage
%%%%%%%%%%%%%%%%%%%%%
%Tables and figures
%%%%%%%%%%%%%%%%%%%%%

\clearpage

\begin{table}[h]
  \caption{Band gaps and band alignments (eV)}
  \label{tbl:Gaps}
  \begin{ruledtabular}
  \begin{tabular}{lccc}
	Parameter & \multicolumn{3}{c}{Structure} \\
    \cline{2-4}
    	& stoichiometric & $\text{I/Pb}=3.22$, $x=0.1$ & $\text{I/Pb}=3.5$, $x=0.2$  \\
	& Fig.~\ref{Fig:Structures}(a) & Fig.~\ref{Fig:Structures}(b) & Fig.~\ref{Fig:Structures}(c) \\
    \hline
	$E_\mathrm{g}$ PBE (no SOC) & 1.47 & 1.59 & 1.83 \\
	$E_\mathrm{g}$ PBE (with SOC) & 0.38 & 0.66 & 1.12 \\
	$\delta_\mathrm{v}$ PBE (with SOC)\textsuperscript{\emph{a}} & n/a & 0.13 & 0.13 \\
	$\delta_\mathrm{c}$ PBE (with SOC)\textsuperscript{\emph{a}} & n/a & 0.15 & 0.61 \\
  \end{tabular} % keep next line empty
  \end{ruledtabular}

	\raggedright
	\textsuperscript{\emph{a}} See Fig.~\ref{Fig:Alignment}(a) for definition of band alignments. The positive value of $\delta_\mathrm{v}$ indicates that the VBE is shifted down in energy relative to the stoichiometric phase. The positive value of $\delta_\mathrm{c}$ corresponds to the upward shift of the CBE in energy.
\end{table}

\begin{table}[h]
  \caption{Effective masses of electron and holes $m^*/m_e$.}
  \label{tbl:Mstar}
  \begin{ruledtabular}
  \begin{tabular}{lccc}
	Carrier type, & \multicolumn{3}{c}{Structure} \\
    \cline{2-4}
    	direction & stoichiometric & $\text{I/Pb}=3.22$, $x=0.1$ & $\text{I/Pb}=3.5$, $x=0.2$  \\
	& Fig.~\ref{Fig:Structures}(a) & Fig.~\ref{Fig:Structures}(b) & Fig.~\ref{Fig:Structures}(c) \\
    \hline
	Electrons \\
	~~$R\rightarrow S$ & 0.08 & 0.11 & 0.11 \\
	~~$R\rightarrow U$ & 0.08 & 0.17 & 0.45 \\
	~~$R\rightarrow T$ & 0.08 & 0.21 & 0.45 \\
    \hline
	Holes \\
	~~$R\rightarrow S$ & $-0.09$ & $-0.13$ & $-0.18$ \\
	~~$R\rightarrow U$ & $-0.09$ & $-0.25$ & $-7.4$ \\
	~~$R\rightarrow T$ & $-0.09$ & $-0.42$ & $-7.4$ \\
  \end{tabular}
  \end{ruledtabular}
\end{table}

\end{document}